\documentclass[]{spie}  

\usepackage{amsmath,amsfonts,amssymb}
\usepackage{graphicx}
\usepackage[colorlinks=true, allcolors=blue]{hyperref}
\usepackage{animate}

\title{Battle of the Predictive Wavefront Controls: Comparing Data and Model-Driven Predictive Control for High Contrast Imaging}

\author[a]{J. Fowler}
\author[a]{M.A.M. Van Kooten}
\author[a]{R. Jensen-Clem}
\affil[a]{Univ. of California, Santa Cruz (United States)}

\authorinfo{Further author information: (Send correspondence to J. Fowler)
E-mail: jumfowle@ucsc.edu}

\begin{document} 
\maketitle

\begin{abstract}
Ground-based high contrast exoplanet imaging requires state-of-the-art adaptive optics (AO) systems in order to detect extremely faint planets next to their brighter host stars. For such extreme AO systems (with high actuator count deformable mirrors over a small field of view), the lag time of the correction (which can impact our system by the amount the wavefront has changed by the time the system is able to apply the correction) which can be anywhere from ~1-5 milliseconds, can cause wavefront errors on spatial scales that lead to speckles at small angular separations from the central star in the final science image. One avenue for correcting these aberrations is predictive control, wherein previous wavefront information is used to predict the future state of the wavefront in one-system-lag's time, and this predicted state is applied as a correction with a deformable mirror. Here, we consider two methods for predictive control: data-driven prediction using empirical orthogonal functions \cite{Guyon2017} and the physically-motivated predictive Fourier control \cite{Poyneer2007}. The performance and robustness of these methods have not previously been compared side-by-side. In this paper, we compare these predictors by applying them as post-facto methods to simulated atmospheres and on-sky telemetry, to investigate the circumstances in which their performance differs, including testing them under different wind speeds, $C_N^2$ profiles, and time lags. We also discuss future plans for testing both algorithms on the Santa Cruz Extreme AO Laboratory (SEAL) testbed \cite{Jensen2021}. This work will inform the next generation of extremely large telescopes (including the European
Extremely Large Telescope, as well as plans for the Giant Magellan Telescope and
the Thirty Meter Telescope), which will depend on predictive control as a key technology to reach the contrasts necessary to directly image rocky planets in the habitable zones of the nearest low-mass stars.

\end{abstract}


\section{INTRODUCTION}
\label{sec:intro}  

For ground-based astronomy, image quality is
impacted by atmospheric turbulence, where varying temperature zones of Earth's atmosphere change the index of refraction in air, which introduces delays in the path length of light travelling from a star to a telescope, leading to aberrations in the final science image. The solution is adaptive optics (AO), wherein phase information is collected with a wavefront sensor and active correction by deformable optics within the telescope path allow these phase delays to be counteracted; hence, wavefront errors are corrected before the light reaches the final science detector. In particular, for an extreme adaptive optics (XAO) system, a bright, on-axis natural guide star and a large number of DM actuators are used to achieve a good correction over a small (e.g. few arcsecond) field of view. 
Extreme AO is required for high contrast imaging systems, which use coronagraphs to null the coherent portion of the starlight. As the wavefront error in an AO system increases, the portion of light that can be rejected by the coronagraph decreases, leading to speckles throughout the image plane that closely mimic planets.

Many current XAO systems (for example the Gemini Planet Imager\cite{Bailey2016} and SPHERE\cite{Milli2017}) see their overall performance limited by temporal effects, termed bandwidth error.  While faster computers and deformable mirrors, as well as increasingly efficient readout
detector technology can shorten an AO system's lag time, often the system lag is dominated
by how long a wavefront sensor needs to expose to get a high signal-to-noise
wavefront measurement. Predictive control can mitigate the bandwidth error without reducing the wavefront sensor exposure time, by
predicting the wavefront of the system one step into the future. While some
atmospheric turbulence will always be stochastic, the bulk motion is accurately represented\cite{Poyneer2009} by the simple Frozen Flow Hypothesis\cite{Taylor1938}, wherein we imagine
static phase screens translating across the telescope pupil at the velocity of their associated wind layers. It is therefore reasonable to conclude that previous measurements of the wavefront contain information that can be used to predict the future shape of the wavefront. 

 With this work, we focus on two disparate predictive control methods: empirical orthogonal functions\cite{Guyon2017} (EOF), as a method that depends
purely on data to find linear relationships through time and space, and predictive
Fourier control\cite{Poyneer2007} (PFC) , which uses data to identify wind layers in a system and inform a
prediction with those layers. As two solutions to the same problem, the performance of EOF and PFC have
yet to be directly compared. With this paper we will study their performance over simulated
atmospheric data and saved on-sky telemetry from the W.M. Keck Observatory. 

In sections \ref{sec:eof} and \ref{sec:pfc}, we introduce empirical orthogonal
functions and predictive Fourier control as two methods of
prediction and present their mathematical formalism. In section \ref{sec:results}, we examine the performance of these
methods, using residual wavefront error, Strehl, and the stability of the
correction as metrics to evaluate their performance, and discuss the comparative performance of the two predictors. In section
\ref{sec:conclusions}, we make conclusions and discuss next steps for testing
these methods on the Santa Cruz Extreme AO Laboratory (SEAL) testbed. 

\section{EMPIRICAL ORTHOGONAL FUNCTIONS AS A DATA-DRIVEN PREDICTOR}
\label{sec:eof}

Empirical Orthogonal Functions (EOF)\cite{Guyon2017} looks for linear relationships in the evolution of the wavefront over time and space. EOF has been successfully run both in simulation\cite{Guyon2017}, during day-time testing at Keck Observatory\cite{Jensen2019}, as well as on-sky at the Keck Observatory\cite{Kooten2022} and the Subaru Telescope\cite{Guyon2018}. Here we will briefly summarize the math behind building and applying the EOF predictive filter; the mathematical formalism is taken from Guyon \& Males (2017)\cite{Guyon2017} unless specified. 

We consider wavefront sensor data at time $t$, to be a collection of m points $\mathbf{w}(t) = [\mathbf{w}_0(t), ... \mathbf{w}_{m-1}(t)]$, that maps to modes (Fourier, Zernike, etc) or zones in the wavefront sensor. We then build history vectors of n of these wavefront sensor exposures each one system time lag apart, such that 

\begin{equation}
\mathbf{h}(t) = 
\begin{bmatrix}
\mathbf{w}_0(t) \\
\mathbf{w}_1(t) \\
... \\
\mathbf{w}_{m-1}(t) \\
\mathbf{w}_0(t -dt) \\
... \\
\mathbf{w}_{m-1}(t-(n-1)dt) 
\end{bmatrix}
\end{equation}

We build a predictive filter $\mathbf{F}$, that when applied to a history vector will predict a future state:

\begin{equation}
    \mathbf{F} \mathbf{h}(t) = 
    \begin{bmatrix}
    \mathbf{w}_0(t+dt) \\
    ... \\
    \mathbf{w}_{m-1}(t+dt)
    \end{bmatrix}
\end{equation}

To calculate $\mathbf{F}$ we build a matrix $\mathbf{D}$ which contains a series of history vectors used for training data, and we compare to a matrix $\mathbf{P}$ which pairs each history vector at time $t$ to its future state at time $t+dt$. From there we calculate the filter $\mathbf{F}$ as a simple minimization:

\begin{equation}
\textrm{min}_\mathbf{F}||\mathbf{D}^T\mathbf{F}^T - \mathbf{P}^T||^2 
\end{equation}

In a departure from Guyon \& Males (2017)\cite{Guyon2017}, we solve this with a least squares pseudo-inversion\cite{Jensen2019}, with a regularization constant $\alpha$ (which we set to $\alpha=1$ for simulations.)

\begin{eqnarray}
\mathbf{F} &=& ((\mathbf{D}^T)^\dagger \mathbf{P}^T)^T \\
\mathbf{F} &=& \mathbf{P}\mathbf{D}^T(\mathbf{D}\mathbf{D}^T + \alpha \mathbf{I})^{-1}
\end{eqnarray}

We found that using 60000 history vectors of training data and history vectors made up of 3 exposures provided a consistent correction (apart from the 7 m/s layer, see further discussion in Section \ref{sec:discuss}), a result also found by Jensen-Clem (2019)\cite{Jensen2019}, and used these parameters for our prediction performance estimates. 

\section{PREDICTIVE FOURIER CONTROL AS A MODEL-DRIVEN PREDICTOR}
\label{sec:pfc}

Predictive Fourier control (PFC)\cite{Poyneer2007} uses data to extract information about wind
layers from the system, and then builds a predictive filter informed by
those wind layers, as an update to a traditional Linear Quadratic Gaussian controller\cite{LGQ1993, Kulcsar2006}. By definition, the predictor starts with no knowledge of the wind speed or direction. Here we will briefly summarize the math behind building and applying the PFC predictive filter; the mathematical formalism is taken from Poyneer (2007)\cite{Poyneer2007} unless specified. (Further details are available in Appendix \ref{sec:pfcmath}.)

In short, PFC consists of 4 main steps:
\begin{enumerate}
    \item Decompose turbulent phase screens into complex spatial Fourier modes. Each Fourier
        mode becomes a separate control problem and a filter is calculated
        individually for each coefficient. For each coefficient:
    \item Make a power spectral density (PSD) function by looking at how a single
        coefficient evolves in temporal frequency space. 
    \item Isolate and fit peaks in the PSDs (which map to wind layers in an atmosphere). 
    \item Feed the coefficients of each fit to a Kalman filter.
\end{enumerate}

From the perspective of a telescope system (under a Frozen Flow assumption\cite{Taylor1938}), wind layers will present as static turbulent scenes 
crossing over the telescope primary mirror at the speed of the wind at that height. Given enough data in time to adequately sample a wind velocity, these pupil crossing times
will have high signal to noise peaks across spatial Fourier modes, as different spatial scales will cross with different frequencies. Therefore, when turbulent phase screens are decomposed into Fourier modes, each coefficient
contains information on a spatial frequency of a particular size that 
maps to a series of non-conflicting controllers. A PSD for each coefficient reveals
the peak frequency, height, and width for each wind layer, which provides input to a controller specific to that coefficient. 

At a given Fourier mode indexed (k, l) (over wavefront sensor subapertures), a peak at some $\nu$ indicates a velocity 
($v_{x}, v_{y}$) component of:
\begin{equation}
    \nu = - \frac{(kv_{x} + lv_{y})}{D}
\end{equation}
where D is the telescope diameter. While in practice, the algorithm is not meant
to deliver identifications of wind-layers in a system, testing the wind
identification on simulated known atmospheres shows recovery of
both single and multi-layer wind layers. Figure \ref{fig:wind_id_good} shows examples of successful recovery of rejected wind layers. See Appendix \ref{sec:pfcmath} for some exploration of less optimal peak recovery and a discussion of potential causes. 

Peaks can be identified and fit using built-in \texttt{scipy} tools (see
Appendix \ref{sec:pfcmath} for further details.) Also present in each PSD is a
peak that occurs at $\nu=0$, due to a 0 Hz frequency peak (DC peak) from the Fourier transform that builds the PSD \cite{Poyneer2007}. DC noise encapsulates the static offset in a system, which may be caused by consistent noise from electric components in a realistic system, or mean offset from zero for any distribution. Wind velocity peaks that occur near zero (in frequency space) are difficult to resolve, as they blend into the static DC peak. Further description of the minimum resolvable peaks and treatment of the DC peak is available in Appendix \ref{sec:pfcmath}. 

For each peak, a height $\sigma^2$, and placement (i.e., peak width and location in angular frequency space) $\alpha$, are fit with
\begin{equation}
    P(\omega) = \frac{\sigma^2}{|1 - \alpha\exp{(-i\omega)}|^2}
\end{equation} 
The complex term $\alpha$ is expressed
$\alpha = |\alpha|e^{i\omega}$, where $\omega$ encapsulates the actual wind
velocity of each layer. (Without the complex phase, $\alpha$ acts as a gain for a typical Linear Quadratic Gaussian controller without extracting and predicting based on wind layers.) 

   \begin{figure} [ht]
   \begin{center}
   \begin{tabular}{c} 
   \includegraphics[width=0.5\linewidth]{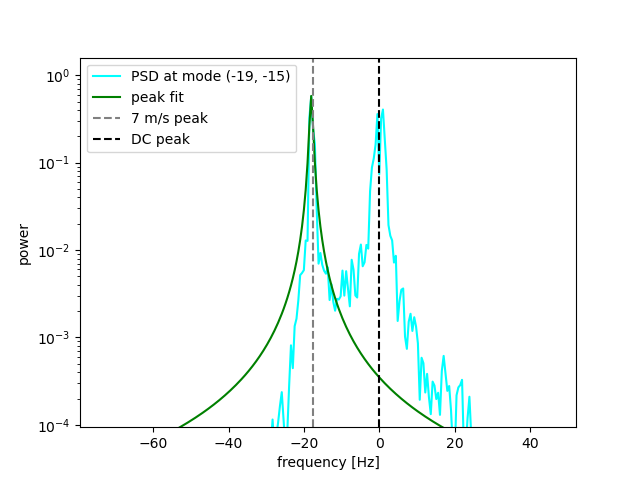}
   \includegraphics[width=0.5\linewidth]{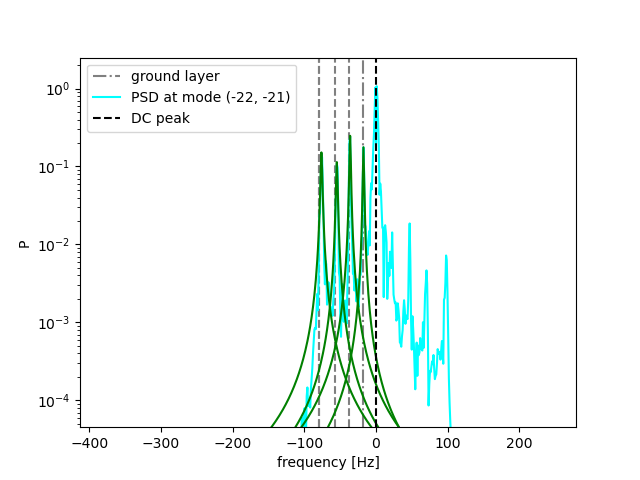}
	\end{tabular}
	\end{center}
   \caption[example] {Automatic peak identification and fitting with \texttt{scipy} tools. Left: Peak identification (for a single layer atmosphere) at the (-19, -15) mode. The 7 m/s peak is correctly identified, along with the DC peak at 0. Right: Peak identification (for a multi-layer atmosphere) at the (-22, -21) mode. All four wind layers are recovered.}
   { \label{fig:wind_id_good} 
}
   \end{figure}

The fits for each $\alpha$ and $\sigma$ inform a Kalman filter, which in turn provides a prediction of
each Fourier coefficient. We predict the state vector of our system 
\begin{equation}
    \mathbf{x}[t] = 
\begin{bmatrix}
   \mathbf{a}(t) & \phi(t+1)  & \phi(t) & \phi(t-1) & d(t-1) & d(t-2)
\end{bmatrix}
\end{equation}
where $\mathbf{a}$ encapsulates the fit parameters, $\phi$ are the wavefront states, and $d$ are commands to the deformable mirror. We predict the next iteration with: 

\begin{equation}
    \mathbf{x}(t) = (\mathbf{I} - \mathbf{K}\mathbf{C})\mathbf{A}\mathbf{x}(t-1) + (\mathbf{I} - \mathbf{K}\mathbf{C})\mathbf{G}d(t-1) + \mathbf{K}y(t)
\end{equation}
where $\mathbf{K}$ are the Kalman gains, $\mathbf{C}$ is a control matrix, $\mathbf{A}$ is a covariance matrix, $\mathbf{G}$ is a matrix to update DM commands, $d(t)$ holds DM commands, and $y(t)$ holds noisy wavefront sensor measurements. For the full definitions and derivations of this Kalman filter see Appendix \ref{sec:pfcmath}, as well as the original treatment \cite{Poyneer2007}. Note that in our implementation, for ease of comparison with EOF, we rewrite the original control law so that we can apply it in open loop. Details of this update to the original form are described in Appendix \ref{sec:pfcmath}.

We found that using 10000 frames of data to build our PSDs and windows between 1024-2048 exposures provided a consistent correction (see Appendix \ref{sec:pfcmath} for further description of the windowing and its impacts), and used these parameters for our prediction performance estimates. 

\section{RESULTS}
\label{sec:results}

\subsection{Simulated Single and Multi-Layer Turbulence}

Using \texttt{hcipy}, the High Contrast Imaging package for Python\cite{hcipy}, we simulated three test case atmospheres: two with single layer atmospheres with a wind speed of 7 m/s (split into $v_x, v_y = 2.13, 6.67$) and 14 m/s (split into $v_x, v_y = 2, 13.5$), and one that included 8 layers according to a C$_N^2$ and wind velocity study of Maunakea weather\cite{KAON303}. The layers and heights from this profile are described in Table \ref{tab:chun}; all layers in the multi-layer simulation had a random direction assigned. For the purposes of this simulation we used an 8 meter primary mirror diameter, an $r_0=15$cm at $500$nm, simulated at a resolution of 48 by 48 pixels (which maps to an idealized wavefront sensor of 48 by 48 pixels), and evolved with a $0.5$ms time step. This simulates a Gemini Planet Imager-like system, which was the instrument for the original PFC simulations\cite{Poyneer2007}. We consider an idealized wavefront sensor whose output is exactly the simulated 48 by 48 pixel phase screen at a given time step. I.e., we use a perfect wavefront sensor (there is no difference between the phase screen and the phase screen we sense) and a perfect deformable mirror (there is no difference between the phase screen we predict and the correction we apply) and focus only on how faithfully we can predict the wavefront given a choice of time delay. 

   \begin{figure} [ht!]
   \begin{center}
   \begin{tabular}{c} 
   \includegraphics[width=0.5\linewidth]{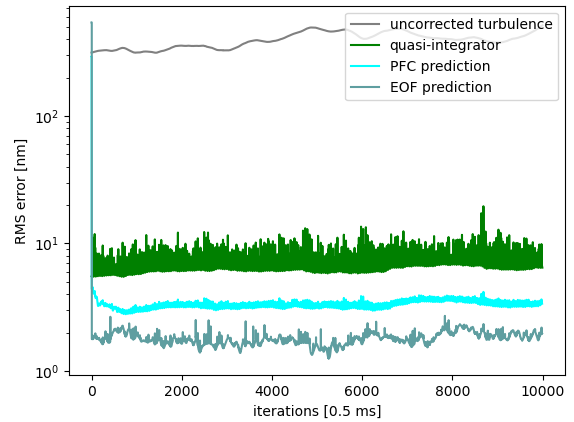}  \includegraphics[width=0.495\linewidth]{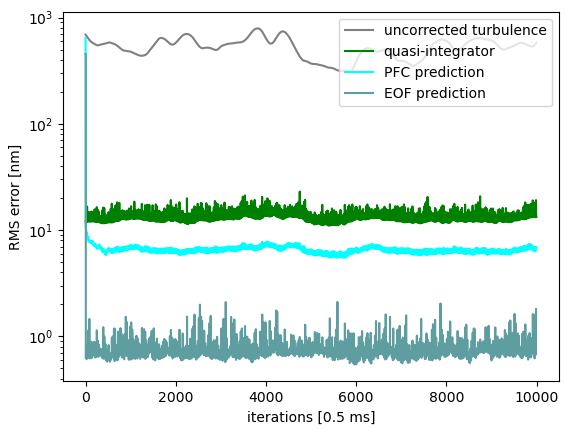}
	\end{tabular}
	\end{center}
   \caption[example] 
   { \label{fig:sim_results} 
Root mean square residuals for the uncorrected turbulence, a quasi-integrator (the known state of the wavefront applied as a correction 2 steps behind), and the two prediction methods. Left: Single 7 m/s wind layer. Right: Single 14 m/s wind layer. Both simulated atmospheres show a performance improvement over a classic integrator from prediction, with EOF performing the best across both atmospheres. } 
   \end{figure} 
   
      \begin{figure} [ht!]
   \begin{center}
   \begin{tabular}{c} 
     \includegraphics[width=0.5\linewidth]{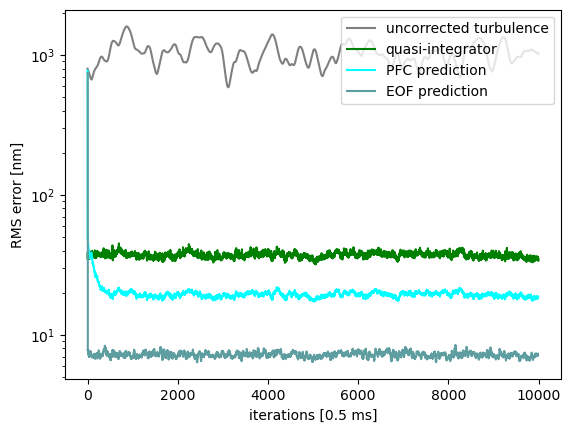}
	\end{tabular}
	\end{center}
   \caption[example] 
   { \label{fig:multi_results} Uncorrected turbulence, quasi-integrator with a two-step lag, and two predictors applied to a simulated 8 layer atmosphere from Neyman (2004)\cite{KAON303}. Both prediction methods show improvement over a classic integrator, and EOF shows the most dramatic performance improvement. }
   \end{figure}

     \begin{figure} [ht]
   \begin{center}
   \begin{tabular}{c} 
   \includegraphics[width=0.56\linewidth]{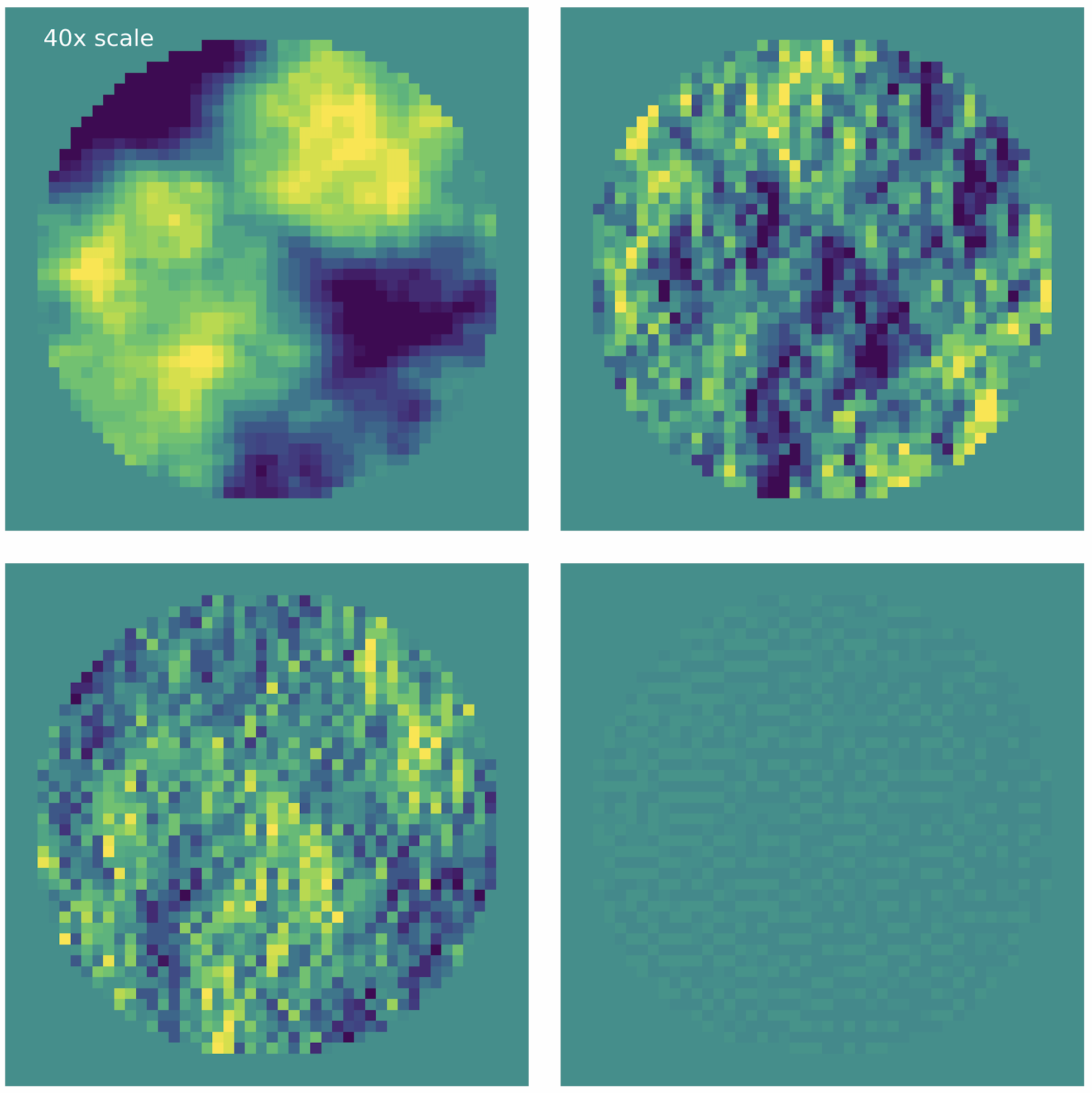}
	\end{tabular}
	\end{center}
   \caption[example] 
   { \label{fig:video_sim} 
(See SPIE Proceeding for video.) Turbulence residuals evolving in real time for the 14 m/s wind simulated atmosphere. Top left: Uncorrected atmospheric turbulence. Top right: Quasi-integrator residuals. Bottom left: PFC predictor residuals. Bottom right: EOF predictor residuals. The open-loop turbulence is at 40x the scale of the other three residuals.}
   \end{figure}

\begin{table}[ht]
\caption{Wind layers simulated based on Neyman (2004)\cite{KAON303}, to emulate nominal performance at Maunakea.} 
\label{tab:chun}
\begin{center}       
\begin{tabular}{|l|l|l|} 
\hline
\rule[-1ex]{0pt}{3.5ex}  Velocity & Height & $C_N^2$  \\
\hline
\rule[-1ex]{0pt}{3.5ex}  6.7 m/s & 0.0 km & 0.369 $\times 10^{-12}$   \\
\hline
\rule[-1ex]{0pt}{3.5ex}  13.9 m/s & 2.1 km & 0.219 $\times 10^{-12}$   \\
\hline
\rule[-1ex]{0pt}{3.5ex}  20.8 m/s & 4.1 km & 0.127 $\times 10^{-12}$   \\
\hline
\rule[-1ex]{0pt}{3.5ex}  29.0 m/s & 6.5 km & 0.101 $\times 10^{-12}$   \\
\hline
\rule[-1ex]{0pt}{3.5ex}  29.0 m/s & 9.0 km & 0.046 $\times 10^{-12}$   \\
\hline
\rule[-1ex]{0pt}{3.5ex}  29.0 m/s & 12.0 km & 0.111 $\times 10^{-12}$   \\
\hline
\rule[-1ex]{0pt}{3.5ex}  29.0 m/s & 14.8 km & 0.027 $\times 10^{-12}$   \\
\hline 
\end{tabular}
\end{center}
\end{table}

We calculated residuals by subtracting the original simulated wavefront from the predicted wavefront in phase, and then calculate the root mean square (RMS) error at $500$nm. For the sake of comparison, we also estimate the performance of a quasi-integrator, where we subtract the simulated phase from itself with a two-step delay (i.e., if the system could apply a perfect correction two steps behind.) For this initial implementation we found that edge effects could have a major impact on the residual error, so as an interim measure we calculated the RMS error only over the inner 7 meter diameter of phase correction. Figure \ref{fig:sim_results} shows the results for the two predictors. For the 7 and 14 m/s single-layer turbulence examples, both predictors show improvement over the quasi-integrator, and EOF shows an advantage between the two predictors. Figure \ref{fig:video_sim} shows a video of turbulence and residuals of the 14 m/s atmosphere evolving over time for the 10000 time step correction. Figure \ref{fig:multi_results} shows the results for the multi-layer example, which again shows that both predictors out-perform the integrator, and an advantage to EOF between the two predictors (see Section \ref{sec:discuss} for further elaboration.) 

\subsection{On-Sky Telemetry}

Our telemetry data is from the Keck II AO system running in closed loop on the night of 04-20-2019, previously published in Jensen-Clem (2019)\cite{Jensen2019}. It consists of 120000 time steps of data (taken at $\sim 1.7$ms intervals\cite{Cetre2018}) in the form of DM commands and near-IR pyramid wavefront sensor\cite{Bond2018} (installed in Keck II as part of the KPIC upgrade\cite{KPIC}) phase measurements in units of volts. Keck II is a 10 meter telescope, with a 21 by 21 actuator deformable mirror (we use the pyramid wavefront sensor's 21 by 21 reconstructed residuals to match.)  

As we see only the residuals and applied commands in this system, we must first reconstruct the pseudo-open loop turbulence in microns with 
\begin{gather}
   \textrm{open loop phase} =  (-1\times\textrm{dm commands} + \textrm{wf residuals})\times0.6 \\
   \textrm{integrator residuals} = (\textrm{wf residuals})\times0.6
\end{gather}
The 0.6 factor accounts for system gain and volt-to-micron conversion\cite{Jensen2019}.

\begin{figure} [ht]
   \begin{center}
   \begin{tabular}{c} 
   \includegraphics[width=0.5\linewidth]{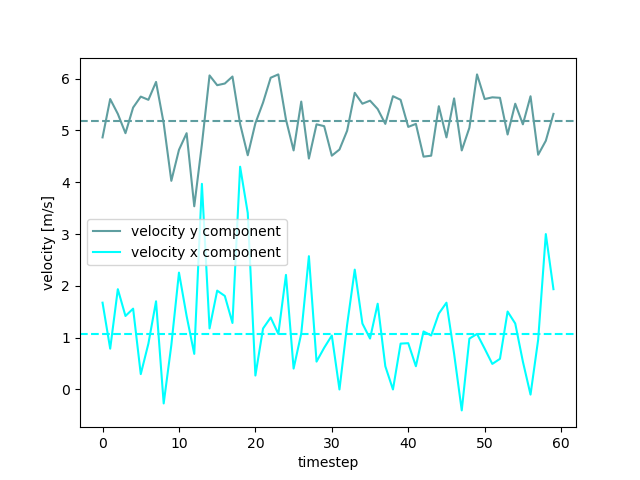}
   \includegraphics[width=0.5\linewidth]{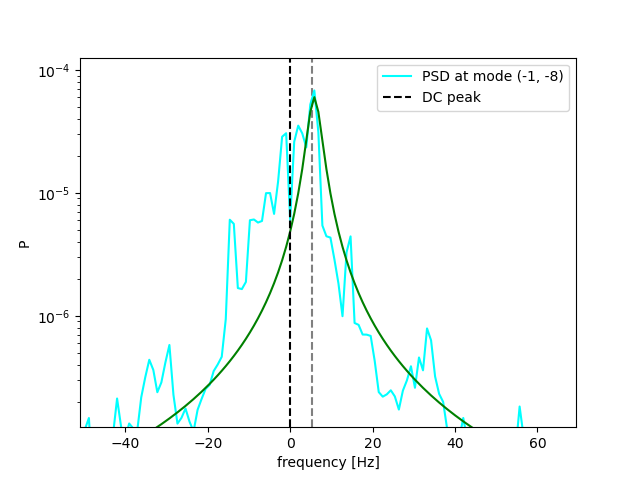}
	\end{tabular}
	\end{center}
   \caption[example] {Left: Canada-France-Hawaii Telescope data from 04-20-19 used to estimate a median x and y velocity component.  Right: Peak identification at the (-1, -8) mode. We identify the ground-layer velocity peak expected from the CFHT data (grey dashed line).}
   { \label{fig:wind_id_telem} 
}
   \end{figure}

Using data from the Maunakea Weather Center\cite{mkwc}, we use the Canada-France-Hawaii Telescope (CFHT) to estimate the wind velocity at the ground layer from the two minutes of Keck telemetry that we use for this analysis. Figure \ref{fig:wind_id_telem} shows the median x and y velocity components from our portion of the night, alongside a peak identification of the ground layer wind velocity from telemetry data. We estimate the ground layer to be $(v_x, v_y) = 1.07, 5.17$m/s. DIMM-reported seeing on that night was $\sim 0.7$", as opposed to MASS-reported seeing of $\sim 0.3$"\cite{mkwc}. Because MASS measures free atmosphere, the discrepancy between the two values implies the turbulence that night was dominated by the ground-layer, making it a good candidate for improvement with predictive control. Using the DIMM value for seeing that night, we find $r_0 \sim 15$cm at $500$nm. 

  \begin{figure} [ht]
   \begin{center}
   \begin{tabular}{c} 

   \includegraphics[width=0.48\linewidth]{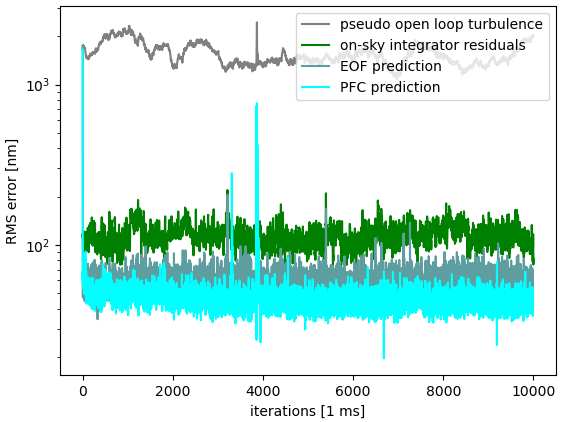}
	\end{tabular}
	\end{center}
   \caption[example] 
   { \label{fig:telem_results} 
 Root mean square residuals for the pseudo-open loop reconstructed turbulence, the true on-sky residuals, and the two prediction methods. Both predictors show improvement over a classic integrator, but PFC shows a modest improvement between the two prediction methods.}
   \end{figure} 
   
     \begin{figure} [ht]
   \begin{center}
   \begin{tabular}{c} 
   \includegraphics[width=0.55\linewidth]{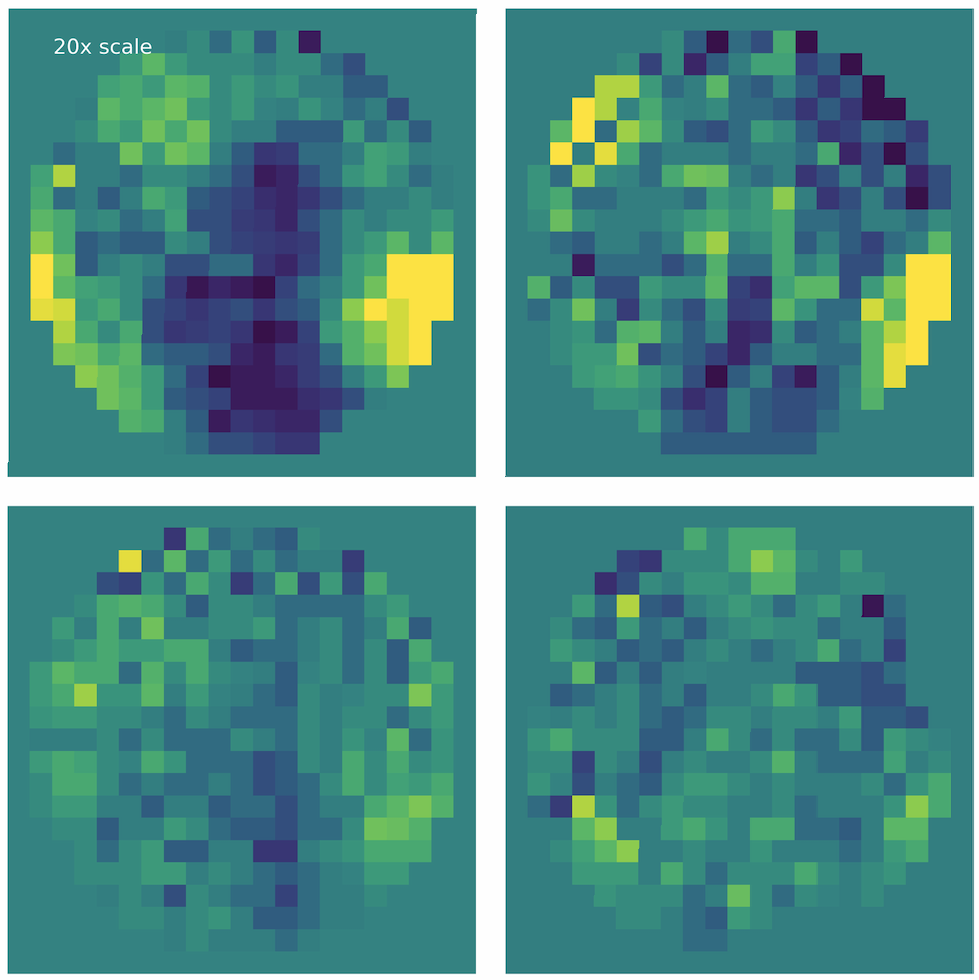}
	\end{tabular}
	\end{center}
   \caption[example] 
   { \label{fig:video_telem} 
(See SPIE Proceeding for video.) Turbulence residuals evolving in real time for the on-sky telemetry. Top left: Pseudo-open loop residuals. Top right: on-sky integrator residuals. Bottom left: PFC predictor residuals. Bottom right: EOF predictor residuals. The open-loop turbulence is at 20x the scale of the other three residuals.}
   \end{figure} 

We applied both predictors to reconstructed open-loop phase screens, calculated residuals by subtracting the predicted screen via each method from the reconstructed screen, and then calculated the root mean square error. Figure \ref{fig:telem_results} shows the results for the two predictors. Both predictors show a $\geq 2$ factor improvement over the on-sky integrator, and in a departure from simulated results,  PFC shows a modest advantage between the two predictors. Figure \ref{fig:video_telem} shows the turbulence residuals evolving over time; note features moving at a consistent velocity are still perceivable in the pseudo-open loop and integrator but give way to whiter noise in the two predictors, which implies our predictor is removing the linear trends imparted by wind layers.

\subsection{Discussion}
\label{sec:discuss}

Table \ref{tab:strehl} compiles the residual wavefront error from each prediction method across each atmosphere and estimates (using the Marechal approximation\cite{hardy1998}) the Strehl ratio achieved by the integrator as well as each predictor. Note that simulations are highly idealized, with no measurement error from the wavefront sensor or fitting error from the deformable mirror.

\begin{table}[ht]
\caption{Estimation of Strehl improvements from residual WFE of each predictor.} 
\label{tab:strehl}
\begin{center}       
\begin{tabular}{|l|l|l|l|l|} 
\hline
\rule[-1ex]{0pt}{3.5ex}  Atmosphere & wavelength & Predictor & WFE & Strehl  \\
\hline
\rule[-1ex]{0pt}{3.5ex}  7 m/s & 500 nm & quasi-integrator & 6.45 +/- 0.9 nm & 92.2 \%  \\
\hline
\rule[-1ex]{0pt}{3.5ex}  7 m/s & 500 nm & PFC predictor & 3.31 +/- 0.2 nm & 95.9 \%  \\
\hline
\rule[-1ex]{0pt}{3.5ex}  7 m/s & 500 nm & EOF predictor & 1.75 +/- 0.2 nm & 97.8 \%  \\
\hline
\rule[-1ex]{0pt}{3.5ex}  14 m/s & 500 nm & quasi-integrator & 12.84 +/- 1.1 nm & 85.1 \%  \\
\hline
\rule[-1ex]{0pt}{3.5ex}  14 m/s & 500 nm & PFC predictor & 6.48 +/- 0.46 nm & 92.2 \%  \\
\hline
\rule[-1ex]{0pt}{3.5ex}  14 m/s & 500 nm & EOF predictor & 0.67 +/- 0.11 nm & 99.2 \%  \\
\hline
\rule[-1ex]{0pt}{3.5ex}  multi-layer & 500 nm & quasi-integrator & 36.84 +/- 1.54 nm & 62.9 \%  \\
\hline
\rule[-1ex]{0pt}{3.5ex}  multi-layer & 500 nm & PFC predictor & 19.39 +/- 2.37 nm & 78.4 \%  \\
\hline
\rule[-1ex]{0pt}{3.5ex}  multi-layer & 500 nm & EOF predictor & 7.19 +/- 0.30 nm & 91.4 \%  \\
\hline
\rule[-1ex]{0pt}{3.5ex}  telemetry & 1633 nm & on-sky integrator & 112.35 +/- 17.27 nm & 64.9 \%  \\
\hline
\rule[-1ex]{0pt}{3.5ex}  telemetry & 1633 nm  & PFC predictor & 44.96 +/- 18.56 nm & 84.1 \%  \\
\hline
\rule[-1ex]{0pt}{3.5ex}  telemetry & 1633 nm  & EOF predictor & 58.43 +/- 8.02 nm & 79.9 \%  \\
\hline
\end{tabular}
\end{center}
\end{table}

For every simulated atmospheric profile and on-sky telemetry we find that both predictors provide an improvement over a classic integrator. PFC provides a consistent improvement of 2-3 over the integrator, while EOF shows more variety. EOF succeeds more notably in picking out one extremely fast wind layer (the single-layer 14 m/s atmosphere), but requires a longer training set (90000 frames of data) to find and remove the 7 m/s wind layer. In contrast, PFC performs consistently with only 10000 frames of training data. Both methods of predictive control also out-perform the integrator in the on-sky telemetry example, which is consistent with the results of Jensen-Clem (2019)\cite{Jensen2019}. For the on-sky telemetry, we find that PFC shows a greater improvement than EOF, in contrast with its performance over the simulated atmospheres. This may be indicative of the way a Kalman filter handles Gaussian noise (a Kalman filter carries the variance of a distribution in its covariance matrix through each iteration\cite{arbook}), as opposed to the treatment of EOF as a purely linear filter. While EOF shows greater performance consistently over atmospheric simulation (which includes no measurement noise), the comparative performance over telemetry may better forecast lab and on-sky performance of the two predictors. However, between the two methods, EOF is less complex to implement, as evinced by the length of Appendix \ref{sec:pfcmath}. 

\section{CONCLUSIONS AND FUTURE WORK}
\label{sec:conclusions}
 
In this paper we have compared an initial implementation of two predictors to examine their relative performance on both simulated data and on-sky telemetry. We find that both the data-driven predictor using empirical orthogonal functions (EOF) and the model-driven predictor using predictive Fourier control (PFC) provide a consistent improvement across every simulated atmospheric profile and the telemetry compared with the integrator. While EOF shows greater improvement than PFC over our three simulated atmospheres, PFC shows greater improvement than EOF over on-sky telemetry data, which may better correlate with future on-sky performance. 

Future work will consist of further optimizing our implementation of these two prediction methods, including focusing on noise estimation and a more robust fit to the DC peak for PFC, and better training data and history vector length optimization for EOF.
Finally, to build on our simulation work, we will implement these predictors as lab experiments on the Santa Cruz Extreme AO Laboratory testbed (SEAL)\cite{Jensen2021}. With this testbed, we will be able to apply realistic turbulence at spatial scales smaller than the resolution of our wavefront sensor (true to the physical nature of atmospheric turbulence) with a spatial light modulator Van Kooten (2022, in prep), correct with with a high-stroke 97 actuator ALPAO and a Boston kilo-DM, and do wavefront sensing with either a Shack-Hartmann or Pyramid wavefront sensor. While simulations are a core initial element of testing the performance of a method, we expect lab demonstrations will unlock key information about the application and performance of these two predictive control methods.

\appendix    

\section{PREDICTIVE FOURIER CONTROL MATH IN DETAIL}
\label{sec:pfcmath}

In this Appendix, we fully describe the mathematical formalism for Predictive Fourier Control\cite{Poyneer2007}, as well as how we have implemented it, and any deviations from the original method. For consistency across EOF and PFC, all matrices are defined in the convention of m rows by n columns, in
departure from typical control engineering conventions and the original formulation from Poyneer (2007)\cite{Poyneer2007}.

\subsection{Decomposition Into Complex Fourier Modes}

We decompose each turbulent scene into complex Fourier modes. Given some mode $(k,l)$, over some x by y pixel grid, each mode is defined by 

\begin{equation}
    f_{k,l}[x,y] = \cos\left(2\pi\frac{kx + ly}{N}\right) + i\sin\left(2\pi\frac{kx + ly}{N}\right)
\end{equation}
where N is the total number of modes.
(Note that the original paper from Poyneer (2007)\cite{Poyneer2007} contains a typo in its casting of each cos/sin term as complex, which we do not duplicate.) 

Fourier modes are meant to operate over square regions, and many telescope pupils are circular; both our Keck telemetry and simulations are over a circular pupil. Starting with a circular pupil embedded within a square grid, and applying a mask for the non-used elements to both the telescope pupil and the Fourier mode allows for a decomposition of the original turbulence image at near machine precision as mathematically proved in Poyneer (2005) \cite{Poyneer2005}, and experimentally reproduced in this work.

\begin{figure} [ht]
   \begin{center}
   \begin{tabular}{c} 
   \includegraphics[width=0.25\linewidth]{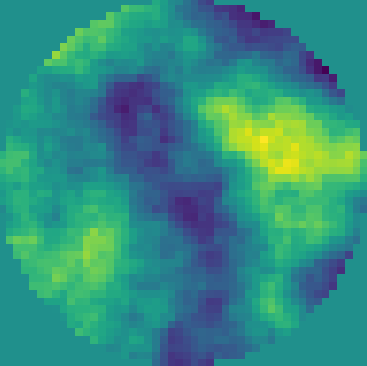}
      \includegraphics[width=0.25\linewidth]{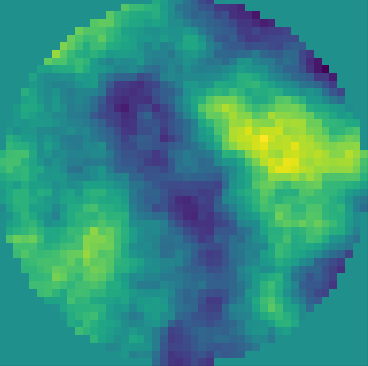}
      \includegraphics[width=0.25\linewidth]{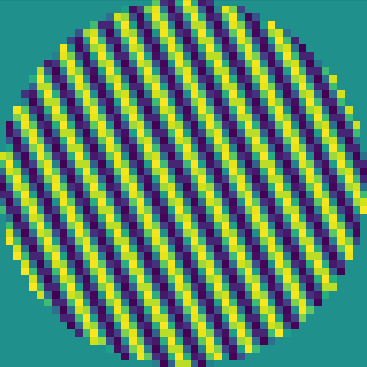} \includegraphics[width=0.25\linewidth]{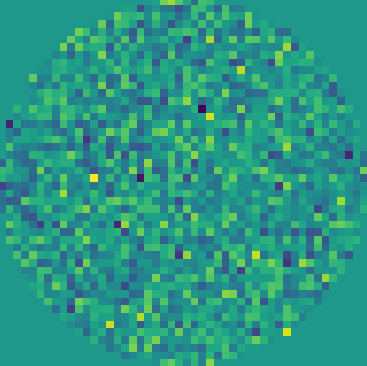}
	\end{tabular}
	\end{center}
   \caption[example] 
   { \label{fig:circular_pupil} 
   Left: Simulated turbulent screen. After Zernike modes are removed from the image, we are left with a circular pupil, and mask the edges within the square. Center left: Reconstruction of the original turbulent scene with the complex Fourier modes, projected into the circular pupil. Center right: Real values of an example complex Fourier mode projected over the same pupil. Right: Residuals between the reconstructed turbulence and the original scene. Residuals are on the order of $1\times10^{-12}$. }
   \end{figure}

Figure \ref{fig:circular_pupil} shows an example of a residual turbulent screen, a reconstruction of the same turbulent screen built with complex Fourier modes, the real component of a complex Fourier mode projected onto the same pupil, and the error introduced during this decomposition, on the order of $1\times10^{-12}$, as compared to the phase of the original turbulence which has a median of $\sim 850$nm (the ratio of which approaches the machine precision of Python.)

\subsection{Creating Power Spectral Densities and Identifying Peaks}

To examine the coefficient for each Fourier mode, we take a periodogram in temporal frequency space. We use a Welch periodogram\cite{statsbook} with a Hamming window, using \texttt{scipy.signal.welch} and \texttt{scipy.signal.get\_window}. 

We use windows of $n=1024$ or $n=2048$, depending on the noise present in the data. Longer windows provide the ability to resolve peaks further from each other and from the 0Hz DC peak, but increase noise in the periodogram. For our smallest window, $n=1024$ frames, our minimum resolvable frequency would be $\nu_{min}=1024\times t_{int}=2-0.6$Hz, for our 0.5-1.7 ms resolutions in simulated data and telemetry. We do not record a peak as input to the Kalman filter unless it is $\nu_{min}$ from 0, as well as other identified peaks. 

\begin{figure} [ht]
   \begin{center}
   \begin{tabular}{c} 
   \includegraphics[width=0.5\linewidth]{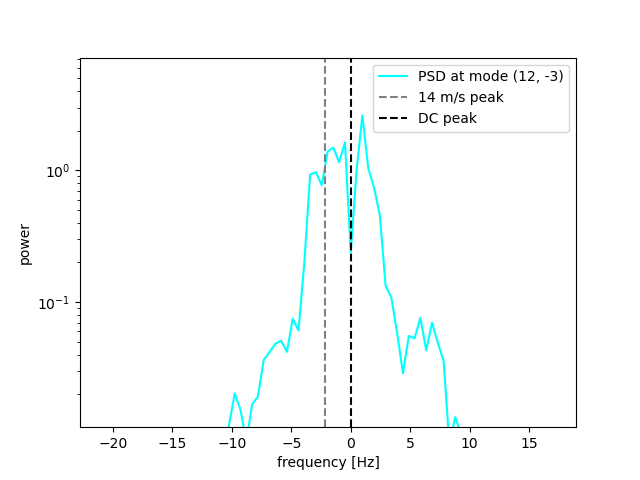}
   \includegraphics[width=0.5\linewidth]{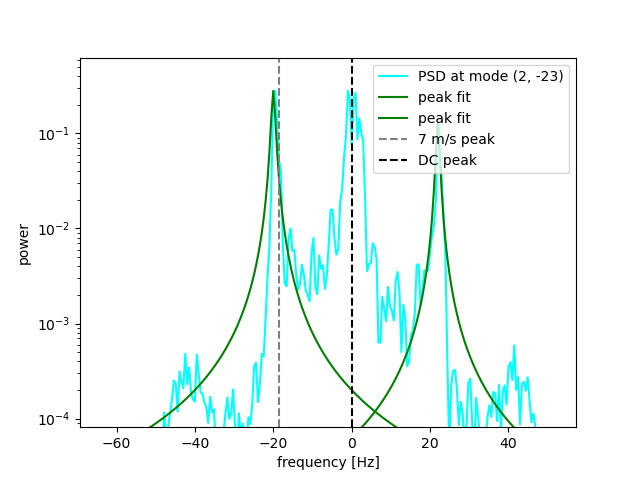}
	\end{tabular}
	\end{center}
   \caption[example] {Left: With the expected wind layer for the (12, -3) coefficient quite close to the DC layer at 0, we are unable to recover the peak from this mode.  Right: Extraneous peak identification at the (-2, -23) mode, as well as the expected 7 m/s peak, likely due to aliasing. Dashed lines in gray and black represent the expected peak placement from the 7 m/s layer and the DC layer respectively. }
   { \label{fig:wind_id_bad} 
}
   \end{figure}

We find the location of each peak in the data with \texttt{scipy.signal.find\_peaks}, and fit it with \texttt{scipy.optimize} \texttt{.curve\_fit}. Peak identification appears repeatable over single layer atmospheres (as shown in Table \ref{tab:peak_id}), though Figure \ref{fig:wind_id_bad} shows some examples of sub-optimal peak identification and fitting given known wind layers, which may be due to spectral leakage or scalloping \cite{arbook}.

\begin{table}[ht]
\caption{Successful (within 2 Hz) peak identifications for varying atmospheric profiles.} 
\label{tab:peak_id}
\begin{center}       
\begin{tabular}{|l|l|l|} 
\hline
\rule[-1ex]{0pt}{3.5ex}  Identified Layer & Correct Peak IDs & Total Modes  \\
\hline
\rule[-1ex]{0pt}{3.5ex}  7 m/s & 1909 & 2304   \\
\hline
\rule[-1ex]{0pt}{3.5ex}  14 m/s & 1989 & 2304   \\
\hline
\rule[-1ex]{0pt}{3.5ex}  telemetry ground layer & 179 & 484   \\
\hline
\end{tabular}
\end{center}
\end{table}

\subsection{Fitting $\alpha$ and $\sigma$}

Each peak can be defined with the power law:

\begin{equation}
    P(\omega) = \frac{\sigma^2}{|1 - \alpha e^{(-i\omega)}|^2}
\end{equation}
where $\sigma^2$ describes the peak height for a given coefficient and complex
$\alpha = |\alpha|e^{i\omega}$ describes the placement (i.e. peak width and velocity) of a given
wind layer. $\nu$ is a natural frequency space in Hz,
whereas the phase $\omega$ describes angular frequency that goes with $\omega =
-2\pi\nu t_{int}$, where $t_{int}$ is the time between each frame of data. 

\subsection{Building the State Vector and Covariance Matrix}

In keeping with the traditional Linear Quadratic Gaussian (LGQ) control, we use a Kalman filter to predict forward the state of the system.
While the original Poyneer (2007) \cite{Poyneer2007} paper describes a close-loop control law, we
reformulate it to run in open loop, adjusting the covariance matrix
$\mathbf{A}$, as well as control matrix $\mathbf{C}$. The original close loop law
includes terms in $\mathbf{A}$ and $\mathbf{C}$ to subtract off a deformable
mirror (DM) command. Our update renders the DM commands in the state vector unnecessary for the Kalman filter (which could impact computational performance), but in this initial implementation
we keep them for simplicity in following the original math. We denote these
changes in each matrix with an$^{*}$. 

The state vector $\mathbf{x}$ is a vector of (L+6) elements that consists of 

\begin{equation}
\mathbf{x}(t) = 
\begin{bmatrix}
   \mathbf{a}(t) & \phi(t+1) & \phi(t) & \phi(t-1) & d(t-1) & d(t-2)
\end{bmatrix}
\end{equation}
where $\phi(t)$ and $d(t)$ map to the wavefront coefficient at each Fourier mode, and
the command to the deformable mirror respectively. The system operates with a two-step time delay, and is centered one step off from the deformable mirror and one step off from the wavefront sensor in opposite directions. At the time of some system measurement $y(t)$ taken at $t=t_0$, the wavefront is at time $t=t_0+dt$, and the deformable mirror has information from time $t=t_0-dt$; this is why we examine $\phi(t+1), \phi(t)$ as compared to $d(t-1), d(t-2)$. 

$\mathbf{a}$ is a vector of (L+1) elements that encapsulates information on every identified wind layer at that
coefficient. 

\begin{gather}
    \mathbf{a} = (a_0, a_1, ..., a_L) \\
    a(t)_L = \alpha_La(t-1) + w(t) 
\end{gather} 
Each $a_L$ evolves according to a first order auto-regressive process, AR(1)\cite{arbook}, driven by the $\alpha_L$ for a given
wind layer and complex white noise $w(t)$. The periodogram for each Fourier mode includes a direct current (DC) layer peak at $\nu=0$, for which we include $\alpha_0=0.999$ and $\sigma^2=\textrm{max}[P(\nu)]$, where $P(\nu)$ is the PSD. (Future implementations could fit the $\sigma^2_{DC}$ term in a way that is more physically relevant.)
    
The state vector $\mathbf{x}$ evolves according to the state equation:
\begin{equation}
    \mathbf{x}(t+1) = \mathbf{A}\mathbf{x}(t) + \mathbf{G}d(t) + \mathbf{B}\mathbf{w}(t) 
\end{equation}

with the (L+6) by (L+6) covariance matrix 
\begin{equation}
\mathbf{A} = 
\begin{bmatrix}
          \mathbf{R} & \mathbf{0}_{L+1 x 1} & \mathbf{0}_{L+1 x 1} & \mathbf{0}_{L+1 x 1} & \mathbf{0}_{L+1 x 1} & \mathbf{0}_{L+1 x 1} \\
          \mathbf{1}_{1 x L+1} & 0 & 0 & 0 & 0 & 0 \\
          \mathbf{0}_{1 x L+1} & 1 & 0 & 0 & 0 & 0 \\
          \mathbf{0}_{1 x L+1} & 0 & 1 & 0 & 0 & 0 \\
          \mathbf{0}_{1 x L+1} & 0 & 0 & 0 & 0 & 0 \\
          \mathbf{0}_{1 x L+1} & 0 & 0 & 0 & 0^{*} & 0 
\end{bmatrix}
\end{equation}

where (L+1) by (L+1) $\mathbf{R}$ holds the $\alpha_L$ values on the diagonals such that 
\begin{equation}
 \mathbf{R} = \textrm{Diag}(\alpha_0, \alpha_1, ..., \alpha_L)
\end{equation}

The 1 by (L+6) DM update matrix is defined as 
\begin{equation}
\mathbf{G} = 
\begin{bmatrix}
    \mathbf{0}_{1 x L+1} & 0 & 0 & 0 & 1 & 0
\end{bmatrix}
\end{equation}

Finally, an (L+6) by (L+1) noise propagator matrix 
\begin{equation}
\mathbf{B} = 
\begin{bmatrix}
    \mathbf{I}_{L+1 x L+1} \\
                  \mathbf{0}_{5 x L+1} 
\end{bmatrix}
\end{equation}

The measurement in the system is defined with
\begin{equation}
    y[t] = \mathbf{C}\mathbf{x}[t]+v[t]
\end{equation}
where $v[t]$ is the measurement noise, which is assumed to be zero-mean Gaussian white noise, and 1 by (L+6) $\mathbf{C}$ is the
control matrix 
\begin{equation}
\mathbf{C} =
\begin{bmatrix} 
    \mathbf{0}_{1 x L+1} &  0 &  0 &  1 & 0^{*} & 0 
\end{bmatrix}
\end{equation}

\subsection{Steady State Solutions with the Algebraic Ricatti Equation}

Instead of recalculating the Kalman filter at each step -- which is quite
computationally expensive, we opt to use the Algebraic Ricatti Equation (ARE)\cite{DARE},
specifically with a discrete solver, as the covariance matrices will reach a
steady-state solution. We use \texttt{scipy.linalg.solve\_discrete\_are} to calculate $\mathbf{P}_s$. Poyneer (2007)\cite{Poyneer2007} outlines this equation as:

\begin{equation}
    \mathbf{P_s} = \mathbf{AP_sA^H} + \mathbf{BP_wB^H} - \mathbf{AP_wC^H}(\mathbf{CP_sC^H} + \mathbf{P_v})^{-1}\mathbf{CP_sA^H} \\
\end{equation}
with $\mathbf{P_v}$ as a scalar variance of the white noise distribution and (L+1) by (L+1)
\begin{equation}
    \mathbf{P_w} = \textrm{Diag}(\sigma^2_0, \sigma^2_1, ..., \sigma^2_L)
\end{equation}
We note that this use of the discrete ARE differs from the typical form in that it uses Hermitian conjugates of the known coefficient matrices $\mathbf{A}\rightarrow\mathbf{A}^H$ and $\mathbf{B}\rightarrow\mathbf{C}^H$, and builds the noise term with $\mathbf{Q}\rightarrow\mathbf{BP_w}\mathbf{B}^H$\cite{DARE, scipy}. 

Finally we calculate the Kalman gains (a vector of L+6 elements) for a given coefficient with:
\begin{equation}
\mathbf{K} = \mathbf{P_sC^H}(\mathbf{CP_sC^H} + \mathbf{P_v})^{-1}
\end{equation}
which builds all of the pieces for the update equation to predict the next step:
\begin{equation}
    \mathbf{x}(t) = (\mathbf{I} - \mathbf{K}\mathbf{C})\mathbf{A}\mathbf{x}(t-1) + (\mathbf{I} - \mathbf{K}\mathbf{C})\mathbf{G}d(t-1) + \mathbf{K}y(t)
\end{equation}

From the newly calculated state vector we pull the $L+1$st element as the prediction of the wavefront $\phi(t+1)$. 

\acknowledgments 
 
Many thanks to Don Gavel, who gave up many days of his retirement to give me a crash course in control theory and help me untangle these fundamental papers,
as well as Lisa Poyneer for advising me on ways to improve my implementation of
Predictive Fourier Control. 

Some of the data presented herein were obtained at the W. M. Keck Observatory, which is operated as a scientific partnership among the California Institute of Technology, the University of California and the National Aeronautics and Space Administration. The Observatory was made possible by the generous financial support of the W. M. Keck Foundation.
The authors wish to recognize and acknowledge the very significant cultural role and reverence that the summit of Maunakea has always had within the indigenous Hawaiian community.  We are most fortunate to have the opportunity to conduct observations from this mountain.

\bibliography{predictive_battle} 
\bibliographystyle{spiebib} 

\end{document}